# Correlation between the Extraordinary Hall Effect and Resistivity.

A. Gerber, A. Milner, A. Finkler, M. Karpovski, L. Goldsmith

Raymond and Beverly Sackler Faculty of Exact Sciences,
School of Physics and Astronomy,
Tel Aviv University
Ramat Aviv 69978 Tel Aviv, Israel

J. Tuaillon-Combes, O. Boisron, P. Mélinon, A. Perez

Département de Physique des Matériaux,
Université Claude Bernard Lyon 1,
F-69622 Villeurbanne, France

We study the contribution of different types of scattering sources to the extraordinary Hall effect. Scattering by magnetic nano-particles embedded in normal-metal matrix, insulating impurities in magnetic matrix, surface scattering and temperature dependent scattering are experimentally tested. Our new data, as well as previously published results on a variety of materials, are fairly interpreted by a simple modification of the skew scattering model.

The anomalous or extraordinary Hall effect (EHE) in magnetic materials has remained a poorly understood phenomenon for more than a century since its discovery. Phenomenology of the effect is straightforward. Hall resistivity $\rho_H$ in magnetic materials is described as: $\rho_H = R_0 B + R_{EHE} \mu_0 M$, where the first term presents an ordinary Hall effect, related to the Lorentz force acting on moving charge carriers, and the second one presents the extraordinary Hall effect with $M$ being the macroscopic magnetization and $R_{EHE}$ the extraordinary Hall effect coefficient. Correlation between the Hall signal and magnetization is well established and has been used for a variety of applications [1, 2]. Problems arise when theoretical models are confronted by experimental data, for example when correlation between the EHE and resistivity is discussed.

The EHE in many cases is much larger than the ordinary Hall effect and is generally believed to originate from a spin-dependent scattering that breaks a spatial symmetry in the trajectory of scattered electrons. An additional contribution of the order of magnitude comparable to that of the ordinary Hall effect and independent of any scattering has been proposed recently [3]. Some models assume the carriers to be magnetic and the scattering centers non-magnetic [4-6], while in others the situation is reversed [7, 8]. Since scattering is responsible both for EHE and longitudinal resistivity, link between two parameters is usually claimed. Two types of scattering events are distinguished in the EHE literature [9]. One is referred to as skew scattering and is characterized by a constant spontaneous angle $\theta_s$ at which the scattered carriers are deflected from their original trajectories.
The predicted [9, 10] correlation between the EHE coefficient and resistivity is: $R_{EHE} = A\rho + B\rho^2$. The second term is frequently neglected and a linear ratio between $R_{EHE}$ and $\rho$ is mentioned. The other scattering mechanism, so-called side jump, is quantum mechanical in nature and results in a constant lateral displacement $\Delta y$ of the charge's trajectory at the point of scattering. For the side jump mechanism $R_{EHE} \propto \rho^2$. Because of the different dependence on resistivity of these mechanisms the EHE is usually attributed to the skew scattering when $\rho$ is small (low temperatures and / or pure metals) and to the side jump when $\rho$ is large (high temperatures, concentrated alloys and disordered materials). Superposition of two mechanisms is presented as:

$$R_{EHE} = a\rho + b\rho^2 \qquad (1)$$

where the first term is believed to relate to the skew scattering and the second to the side jump mechanism although it might contain a contribution of the skew scattering as well. A simplified alternative form of presentation is $R_{EHE} = \alpha \rho^n$ with $n = 1$ corresponding to skew scattering, $n = 2$ to the side jump, and intermediate values $1 < n < 2$ are accepted as a superposition of two mechanisms.

Unfortunately, much of experimental data fall far from theoretical expectations; most notorious are cases in which the power index $n$ is found to exceed 2. It has been recently argued [11-13] that in heterogeneous systems, where the mean free path is comparable or greater than the topological modulation length, the simple scaling relationship between $R_{EHE}$ and $\rho$ no longer holds. The Hall resistivity in heterogeneous systems depends on the ratio of relaxation times (mean free paths) in magnetic and nonmagnetic regions, and as a result the power $n$ may be smaller or greater than 2. These arguments have been used to justify higher than 2 power law values found in e.g. Fe/Cr multilayers [14] ($n = 2.6$) and granular films of Co-Ag [15] (n = 3.7). However, significant discrepancies, including $n > 2$, have been found much earlier not only in heterogeneous but also in bulk homogeneous systems as well [16-18].

We, therefore, propose to reconsider the very correlation between the EHE and resistivity. The present work is an attempt to abandon the traditional link between the *total values* of two parameters. Instead, we decompose both EHE and resistivity to contributions generated by different scattering sources and follow the correlation for each source independently.

Let us start with a simple modification of the skew scattering model. Let us assume that only a certain type of scattering events gives rise to skew scattering, the rest do not break the scattering symmetry. We shall call the sources generating skew scattering as "skew" and the rest as "ballast". Let us also assume that the total resistivity $\rho$ follows the Mattheisen's rule $\rho = \rho_0 + \rho_s$, where $\rho_s$ is the contribution of "skew" sources and $\rho_0$ is due to the rest of "ballast" scattering events. Justification of this assumption will be discussed later. $\rho_s$ and $\rho_0$ can be further subdivided if more than two sources are involved. We consider the system in high applied magnetic field with all magnetic moments saturated and aligned along the field. The EHE resistivity in this saturated state is field independent, and we denote it as $\rho_{EHE}$. Transverse current density $J_\perp$ generated by electrons deflected by skew scattering is proportional to the volume density of "skew" centers $n_s$: $J_\perp = \alpha n_s J$, where $J$ is the longitudinal current density. Coefficient $\alpha$ is proportional to the skew angle $\theta_s$. Transverse electric field $E_\perp$ is: $E_\perp = J_\perp \rho = \alpha n_s J \rho = \alpha n_s J (\rho_0 + \rho_s)$, and Hall resistivity $\rho_{EHE}$ is thus given by:

$$\rho_{EHE} = E_\perp / J = \alpha n_s (\rho_0 + \rho_s) \qquad (2)$$

If $\rho_s \propto n_s$, Eq. (2) can be rewritten as:

$$\rho_{EHE} = \gamma \rho_0 \rho_s + \gamma \rho_s^2 \qquad (3)$$

where $\gamma$ is coefficient.

Equations (2) and (3) allow us to analyze the correlation between the measured Hall resistivity and scattering components by varying only one source at time. If $\rho_s$ is kept constant, $\rho_{EHE}$ is expected to be a linear function of $\rho_0$ with a slope proportional to $\rho_s$ and remnant value $\gamma \rho_s^2$ at $\rho_0 = 0$. If $\rho_0$ is kept constant and the skew scattering term $\rho_s$ is varied, $\rho_{EHE}$ becomes a sum of linear and quadratic terms of $\rho_s$ with the coefficient of the linear term proportional to $\rho_0$. Contrary to Eq.1, both linear and quadratic terms originate from the same skew scattering mechanism only.

In the following we present several experiments in which different scattering mechanisms have been varied in a controllable way.

1. Magnetic scattering centers.

Correlation between the EHE resistivity and density of magnetic scattering centers has been studied in a series of dilute planar arrays of Co nano-clusters embedded in Pt matrix. Samples were produced by the low energy clusters beam deposition (LECBD) technique [19, 20]. Co clusters are crystalline in FCC-phase with a narrow distribution of diameters about 3 nm. Under- and over-layers of Pt films of 5 and 15 nm, respectively, were deposited from an electron gun evaporator mounted in the same deposition chamber. The mean thickness of Co clusters, defined as a total deposited mass divided by density of Co, varied by two orders of magnitude between 0.01 and 1.1 nm. An average distance between centers of the nearest clusters is estimated to vary between 37 to 3.5 nm respectively.

The films are nanocrystalline, their overall resistivity is mainly due to boundary scattering. Substitution of Pt crystallites by Co nano-clusters has no visible effect on the total resistivity of the entire series, which is of the order of 40 μΩcm at room temperature. On the other hand, the Hall resistivity depends strongly on the concentration of Co clusters. Typical variation of the Hall resistivity with applied magnetic field is shown in Fig.1 for three samples with effective Co thickness of 0.1, 0.5 and 1 nm, as measured at room temperature. The samples are superparamagnetic with the blocking temperature at about 40 K. Hysteresis is developed in $\rho_H(B)$ curve below this temperature. $\rho_{EHE}$, the saturated extraordinary Hall resistivity, can be found by extrapolating the high field linear slope of $\rho_H(B)$ to zero field. Fig.2 presents the EHE resistivity plotted as a function of Co-clusters planar density $n_s$. $\rho_{EHE}$ varies linearly with $n_s$ in agreement with Eq.(2) for $\rho = \rho_0 + \rho_s = const$.

2. Insulating non-magnetic impurities.

The case of insulating non-magnetic impurities has been tested by adding silica into nickel. Series of Ni-SiO$_2$ films were prepared by co-deposition of Ni and SiO$_2$ in a two-gun e-beam deposition chamber. More details on fabrication of this type of materials have been reported elsewhere [21]. Morphology of disordered mixtures, like Ni-SiO$_2$, changes dramatically as a function of SiO$_2$ concentration. The size of silica clusters increases, fractal structure is developed and, finally, the percolation threshold is reached. In the present experiment we tried to avoid these complications and limited the concentration of SiO$_2$ to a few volume percents only, such that Ni matrix was kept far above the percolation threshold. Resistivity generated by SiO$_2$ inclusions has been defined as: $\rho_{SiO_2} = \rho - \rho_{Ni}$, where $\rho$ is resistivity of a given sample and $\rho_{Ni}$ is resistivity of a pure Ni sample prepared in the same deposition conditions and measured at the same temperature. The contribution of SiO$_2$ impurities to the EHE resistivity is calculated in a similar way as: $\rho_{EHE,SiO_2} = \rho_{EHE} - \rho_{EHE,Ni}$ with the same meaning of indices. $\rho_{EHE,SiO_2}$ is plotted as a function of $\rho_{SiO_2}$ in Fig.3 for 77 K and 290 K. The dependence is independent of temperature and linear, in agreement with Eqs. (2, 3) for the case when the "ballast" resistivity is varied and the "skew" subsystem is kept constant.

3. Surface scattering.

The effect of surface scattering on EHE has been studied [22] in series of thin Ni films with thickness of the order of electronic mean free path. Following the Fuchs-Sondheimer [23] size effect model, external surfaces impose a boundary condition on the electron-distribution function, which enhances the intrinsic, thickness independent bulk resistivity $\rho_b$ to a thickness-dependent resistivity $\rho$. The total longitudinal resistivity $\rho$ and the extraordinary Hall resistivity $\rho_{EHE}$ of a typical series of Ni films, measured at room temperature, is plotted in Fig.4 as a function of film's thickness. Both resistivities are constant in samples thicker than 100 nm and the latter is taken as the bulk value. The surface scattering term can be extracted explicitly as $\rho_{ss} = \rho - \rho_b$. In a similar way, the contribution of surface scattering to the EHE resistivity can be found as $\rho_{EHEss} = \rho_{EHE} - \rho_{EHEb}$, where $\rho_{EHE}$ and $\rho_{EHEb}$ are the EHE resistivity of a given film and bulk respectively. Fig.5 presents $\rho_{EHEss}$ as a function of $\rho_{ss}$ for two sets of Ni films prepared under different deposition conditions (resistivity of the thick sample in series (b) is about three times higher than that in series (a)). Series (a) has been measured at three temperatures: 4.2 K, 77 K and 290 K, and series (b) at 77 K and 290 K. The dependence is linear and independent of temperature for both series. Similar to the case of insulating impurities, the result is consistent with the situation in which "skew" subsystem is kept constant and "ballast" resistivity is modified.

4. Temperature dependent scattering.

The temperature dependence of resistivity is shown in Fig. 6a for the Co-Pt sample with a mean Co thickness of 0.05 nm between 1.5 K and room temperature. The extraordinary Hall effect coefficient $R_{EHE}(T)$ is plotted in Fig. 6b in the same temperature range. The latter has been calculated as: $R_{EHE}(T) = \rho_{EHE}(T)/\mu_0 M_{sat}(T)$, where $M_{sat}(T)$ is the saturated high field magnetization measured by VSM magnetometer and reported in Ref. [20] for a much thicker sample prepared by the same technique as ours. Both $\rho$ and $R_{EHE}$ behave similarly as a function of temperature: they saturate to the remnant value at low temperatures and increase gradually when the sample is heated.

Prior to focusing on the extracted temperature-dependent components of the EHE and longitudinal resistivity, it is illuminating to view a traditional presentation of the total $R_{EHE}$ as a function of the total $\rho$ in linear and log-log plots when temperature is varied. The EHE coefficient $R_{EHE}$ and not $\rho_{EHE}$ will be discussed to filter the temperature change of the saturation magnetization. In addition to our results with Ni and Co-Pt samples, we reproduce several sets of data reported earlier for a range of magnetic materials. These include thin films of iron (19 and 75 nm thick) [24], polycrystalline iron films [25], Ni films [26], sputtered Pt/Au/Co/Pt sandwiches [27], Fe/Cr multilayers with variable interfacial roughness [28], textured Fe/Cr multilayers grown by electron beam evaporation [14], Co/Cu superlattices [29], and Fe-Ag granular alloys [30]. Two latter systems demonstrate large magnetoresistance and we refer to their resistivity in the saturated high field state. $R_{EHE}$ of all the mentioned materials, normalized by their maximal values at the highest reported temperatures (room temperature in most cases) is plotted in Fig.7a as a function of the respective resistivity. The same data in log-log scale is shown in Fig.7b with resistivity of each sample normalized by its highest value. Distribution of slopes in the latter plot, identified with power indices $n$, is disturbingly wide: from 0.8 in one of Pt/Au/Co/Pt sandwiches [27] to 2.6 in Fe/Cr multilayers [14].

The temperature-dependent components $\rho_{th}$ and $R_{EHE,th}$ have been extracted by subtracting the respective remnant values at the lowest measured temperatures. The resulting $R_{EHE,th}$ normalized by their maximal values at the highest reported temperature is plotted in Fig.8 as a function of the respectively normalized $\rho_{th}$. All sets of data seem to collapse along a single straight line, which means that for all these materials the ratio between the temperature dependent components of the extraordinary Hall coefficient and resistivity is close to be constant. It should be noted that this result is not a trivial consequence of a possible smallness of e.g. $R_{EHE,th}$ as compared with $R_{EHE}(T=0)$. In fact, $R_{EHE}$ varies significantly with temperature (see Fig. 7a), the ratio between the helium and room temperature values is about 0.3 in e.g. Fe [24] and Co-Cu superlattices [29], 0.5 in Fe-Ag granular alloys [30] and our Co-Pt arrays; and 0.7 in Pt/Au/Co/Pt sandwiches [27]. The span of resistivity is also wide (Fig.7a): resistivity of Ni [26] increases from about 4 to 12 $\mu\Omega$cm, Co-Pt array from 25 to 40 $\mu\Omega$cm, and 19 nm thick Fe film [24] from 70 to 130 $\mu\Omega$cm.

Seemingly universal linear variation of $R_{EHE,th}$ with $\rho_{th}$ manifested in Fig.8 should be taken with caution. In majority of cases, $R_{EHE,th}$ is fitted better by a two term expression:

$$R_{EHE,th} = a\rho_{th} + b\rho_{th}^2 \qquad (4)$$

where the absolute values of $b$ are much smaller than $a$. If, alternatively, $R_{EHE,th}$ is presented as $R_{EHE,th} = c\rho_{th}^n$, the power index $n$ varies between $n = 0.9$ in Fe-Ag granular alloys [30], to $n = 1.2$ in Co/Cu superlattices [29] and iron films [24], which can be interpreted as a dominance of the linear term $a\rho_{th}$. In some cases deviation of $R_{EHE,th}$ vs $\rho_{th}$ from linearity is significant, e.g. in Co/Pt superlattices reported by Canedy *et al* [31], and the quadratic term $b\rho_{th}^2$ can not be neglected. In the framework of our model all the temperature-dependent data are consistent with the situation in which the remnant ($T = 0$) value of $R_{EHE}$ and an almost linear slope of $R_{EHE,th}$ vs $\rho_{th}$ curve is determined mainly by the core "skew" subsystem and the deviation from the linearity is due to a relatively small skew scattering contribution of the thermal disorder.

Both phonons and thermal spin disorder have been mentioned as possible sources of the EHE. Following Kagan and Maksimov [32] the phonon contribution is expected to be negligible as compared with that of magnons. However, our data provide no evidence for the role of magnons. The latter are expected to be suppressed by high magnetic field at least at low temperatures (T < 15 K for B > 15 Tesla). In our experiments both Hall and longitudinal resistivity were measured up to 16.5 T and $R_{EHE}$ was extrapolated from this high field range. No change in behavior is marked when temperature is raised from 1.5K to room temperature. We are therefore, inclined to believe that the observed modest temperature-dependent contribution to the EHE coefficient is due to phonons.

Temperature dependence of EHE in magnetic granular alloys has been recently treated by Granovsky et al [33]. Correlation of the type (Eq.4) has been predicted at high temperatures only above the Debye temperature, where resistivity is expected to vary linearly with temperature. It should be noted that we find a linear correlation between $R_{EHE,th}$ and $\rho_{th}$ in the entire measured temperature range, including the low temperature limit where resistivity saturates to its remnant value (see Fig. 6a).

Few words need be added to justify our use of Mattheisen's rule. This phenomenological rule is widely accepted as a useful approximation by which the resistivity of metal can be presented as the sum of a temperature-independent residual resistivity (due to defects) and a part due to phonon scattering. This assumption is valid if the two scattering mechanisms operate independently, that is the scattering by imperfections is temperature independent and there are insufficient imperfections to affect significantly the phonon scattering. The rule can be further subdivided if there is more than one type of imperfection (e.g. grain boundaries and surfaces, [23, 34]). Deviations from Mattheisen's rule due to the interference terms are suppressed at high magnetic fields [35], which is the regime of our interest. In heterogeneous magnetic systems, demonstrating the so-called giant magnetoresistance (GMR) effect, Mattheisen's rule is replaced by the two-current model representing a parallel flow of electrons with spins up and down. However, at high magnetic fields when magnetic moments of the system are aligned, the resistivity is given by: $1/\rho = 1/\rho_\uparrow + 1/\rho_\downarrow$, where $\rho_\uparrow$ and $\rho_\downarrow$ are resistivity of electrons with spins up and down respectively. Large GMR effect is due to a large inequality of $\rho_\uparrow$ and $\rho_\downarrow$, and the high field resistivity can be roughly approximated as due to one (lowest) component only: $\rho \approx \rho_\downarrow$. Therefore, Mattheisen's rule can be considered as valid in the high field limit with only one type of carriers left.

Analysis proposed here might help to resolve several puzzles left by the traditional treatment of experimental data. We shall mention just few cases. Caulet et al [27] studied the extraordinary Hall effect in Pt/Au/Co/Pt sandwiches with variable width of Au layer. The authors tested the validity of Eq.1 and noticed several unexpected features: i) despite the high resistivity of their samples the "skew scattering" contribution $a\rho$ was always dominant; ii) while coefficient $a$ was found to increase slightly with thickness of gold layer, coefficient $b$ decreased strongly and even changed sign. We have reexamined the same data by separating the remnant and the temperature dependent components. The temperature dependent terms of all three samples have been found to collapse on a single curve following Eq.4 with $a = 0.95 \pm 0.05$ and $b = 0.08 \pm 0.05$. The difference between the samples is in their remnant low temperature resistivity and not in the temperature dependent terms.

The effect of interfacial roughness on EHE has been studied by Korevinski et al. [28] in a series of Fe-Cr multilayers. Experimental data has been collected as a function of temperature and approximated by $R_{EHE} = \alpha \rho^n$. For the "smoothest" sample one obtained n ≈ 2.0, while for the "roughest" sample n ≈ 2.3. The authors concluded that: $R_{EHE} \propto \rho^2$ relationship is not unique, larger n corresponds to increasing roughness and, therefore larger roughness leads to stronger temperature dependence. Decomposition of the same data to the remnant and temperature-dependent terms leads to a different conclusion: the temperature dependence is identical for all samples. There is no dependence of the thermal component on roughness, which is consistent with our conclusions on the surface scattering component in Ni films (see Fig.5).

In one of the first and widely cited works on transport properties of granular ferromagnets, P. Xiong et al [15] reported a surprisingly high index n = 3.7 in a power law correlation $\rho_{EHE} = \alpha \rho^n$ between the Hall coefficient and longitudinal resistivity of a granular Co-Ag system. The data was accumulated by thermal treatment of samples with a constant volume concentration of Co at different annealing temperatures. Annealing affects the system in many ways: Co crystallizes, grains coalesce, their size increases and density of clusters decreases respectively. Simultaneously, dislocations in matrix are healed and resistivity decreases. Nevertheless, following the common tradition none of these details have been treated separately and only the overall final resistivity has been correlated with the total Hall effect. The power index 3.7 emerged and stimulated new theoretical efforts [13]. Accurate separation of parameters in the framework of our model is impossible in this experiment; hoverer the overall interpretation might be quite simple. An average size of Co clusters has been reported to grow with annealing from 2 nm to 13 nm. The volume density of Co clusters, has, therefore, reduced roughly by a factor of 300, which is consistent with the observed reduction of $\rho_{EHE}$, uncorrelated to the change of resistivity.

It is an almost general perception that low resistivity systems can be treated by the skew scattering model; whereas the side jump model must be applied in any other case. One of many examples is another influential theoretical work by Shufeng Zhang [11] which concentrates on the side jump as the main source for the EHE in multilayered structures because their resistivity are usually much larger than that of individual bulk materials. It seems however, that the total resistivity might be an erroneous parameter, in particular when its significant part is contributed by scattering with negligible spin-orbit interaction. We were successful in analyzing much of experimental data using the skew scattering mechanism only. The analysis was successful also in high resistivity systems, where the side jump mechanism is automatically taken as the only dominating source of the EHE.

To summarize, we have abandoned the traditional comparison of the total values of the extraordinary Hall resistivity and longitudinal resistivity. Instead, we analyzed the correlation between the two parameters by decomposing them to contributions generated by different scattering sources. Two types of scattering sources have been distinguished: (i) "skew" sources that give rise to skew scattering, and (ii) "ballast" sources that do not generate skew scattering by themselves but contribute linearly to the EHE when additional "skew" sources are present. The extraordinary Hall effect is obtained as a combination of resistivity terms of both types of sources. Insulating impurities and surfaces are identified as "ballast" scattering sources. The temperature-dependent contribution, probably that of phonons, can be considered as close to be "ballast" with a relatively small self-skew scattering. All data discussed, measured in a variety of magnetic materials, both new and previously published, can be fairly interpreted in the framework of the proposed modified skew scattering model without involving the quantum side jump mechanism.

We acknowledge stimulating discussions with A. Palevski, R. Mints and Y. Korenblit. This work has been supported in part by AFIRST grant No. 9841 and by the Israel Science Foundation grant No. 220/02.

# References.

**Figure Captions.**

Fig.1. Hall resistivity of three planar arrays of Co nano-clusters embedded in Pt matrix as a function of applied magnetic field. Mean thickness of Co is 0.1 nm (squares), 0.5 nm (stars) and 1 nm (circles). T = 290 K.

Fig.2 The saturated EHE resistivity of planar arrays of Co nano-clusters embedded in Pt matrix as a function of Co-clusters planar density. Solid line is the guide for eyes.

Fig.3. Contribution of $SiO_2$ impurities to the EHE resistivity of a series of Ni-$SiO_2$ films as a function the respective contribution to longitudinal resistivity. T = 290K - open circles, and T = 77 K – solid circles. Solid line is the guide for eyes.

Fig.4. Longitudinal ($\rho$) and EHE ($\rho_{EHE}$) resistivity of a series of thin Ni films as a function of their thickness. T = 290 K.

Fig.5. Contribution of the surface scattering to the EHE resistivity of thin Ni films as a function of the respective contribution to longitudinal resistivity. Series (a) has been measured at three temperatures: 4.2 K, 77 K and 290 K, and series (b) at 77 K and 290 K. Solid lines are the guides for eyes.

Fig.6. Temperature dependence of resistivity (a), and the EHE coefficient (b) of the Co-Pt array sample with the mean Co thickness of 0.05 nm.

Fig.7. Normalized values of the total EHE coefficient $R_{EHE}$ as a function of the total resistivity $\rho$ (a); and as a function of the respectively normalized total resistivity $\rho$ in log-log scale (b), measured at different temperatures. Symbols indicate:

● - Co-Pt arrays;  ◆ - Fe-Cr multilayers [5];  ☆ - Fe-Ag granular alloys [19]; ◇ , ◈ , ⊗ - three Pt/Au/Co/Pt sandwiches [16];  ■ ,  ○ - Fe films (19 and 75 nm thick) [13];  △ -Ni films [15];  □ - Co-Cu superlattice [18];  ⊖ - polycrystalline Fe films [14];  ⊕ -Fe-Cr multilayers [17].

Fig.8. Normalized values of the temperature-dependent component of the EHE coefficients $R_{EHE,th}$ as a function of the respectively normalized values of the temperature-dependent term of resistivity $\rho_{th}$. Symbols indicate the same selection of materials as in Fig.7.

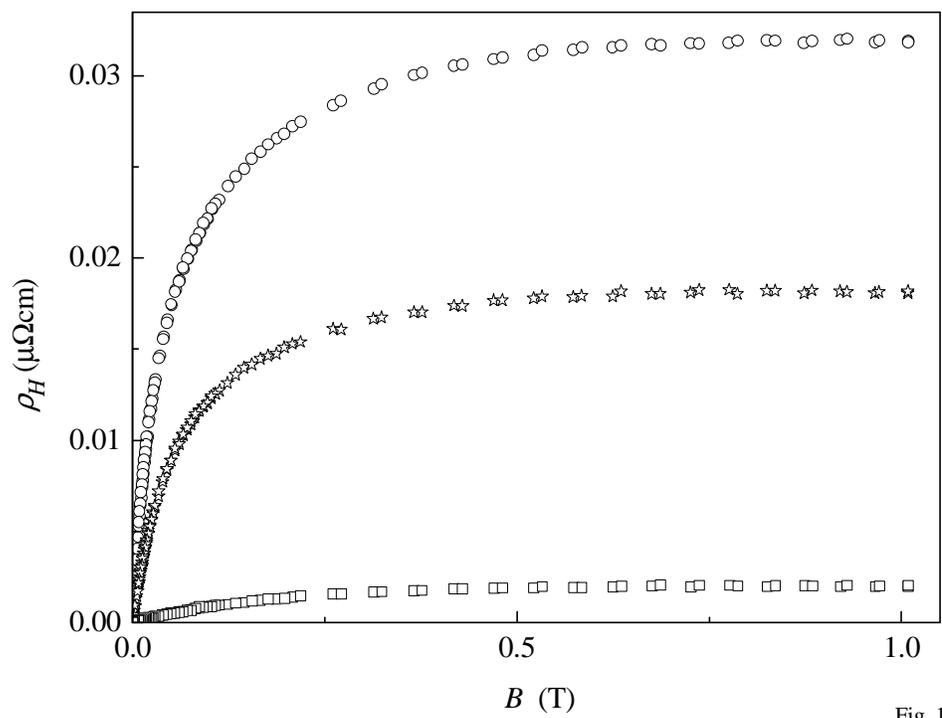

Fig. 1

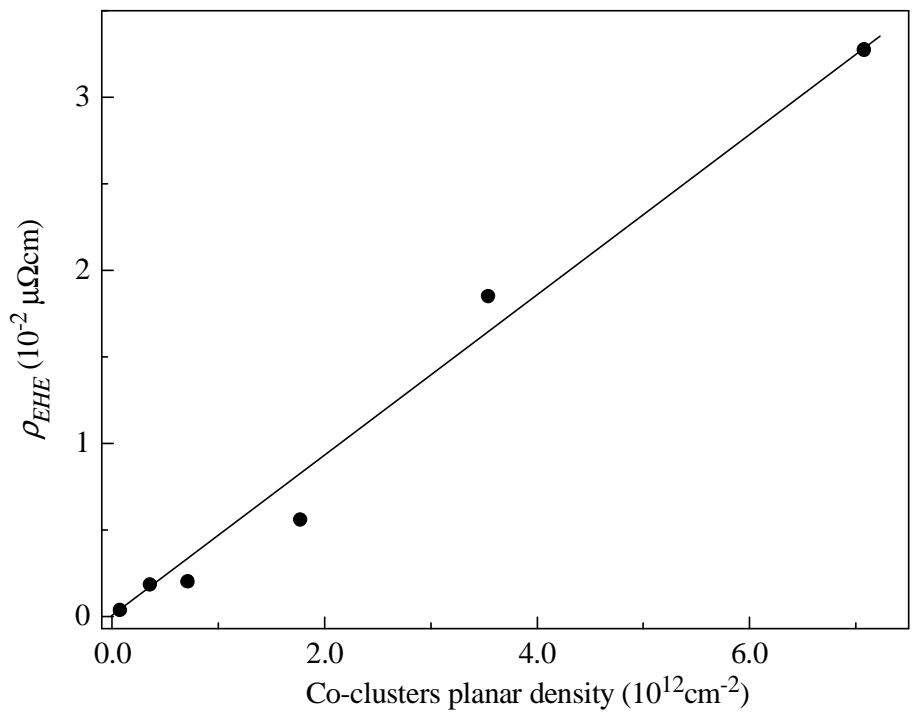

Fig.2

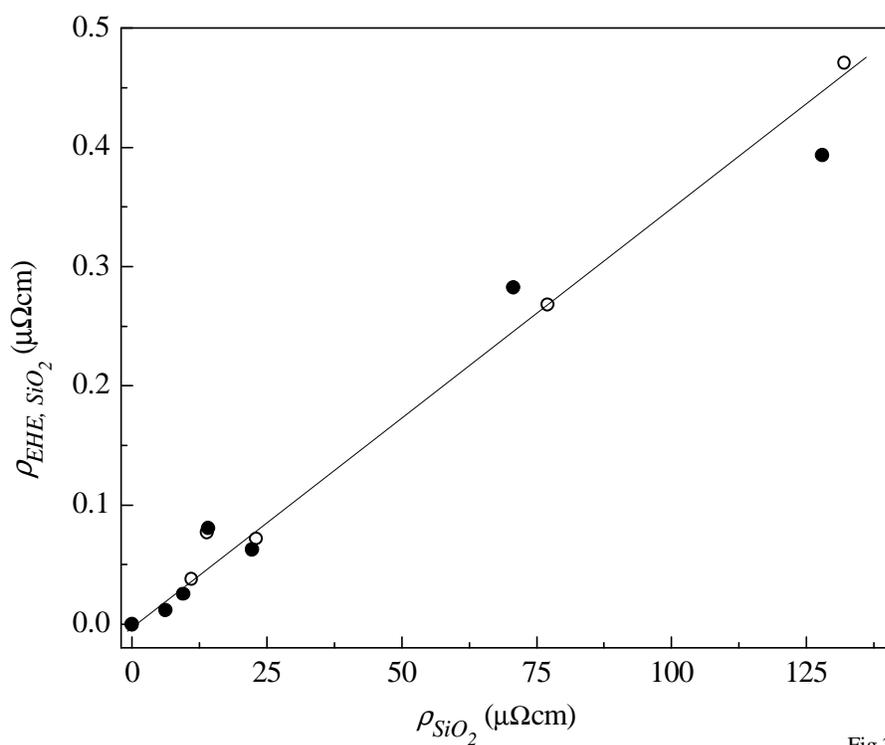

Fig.3

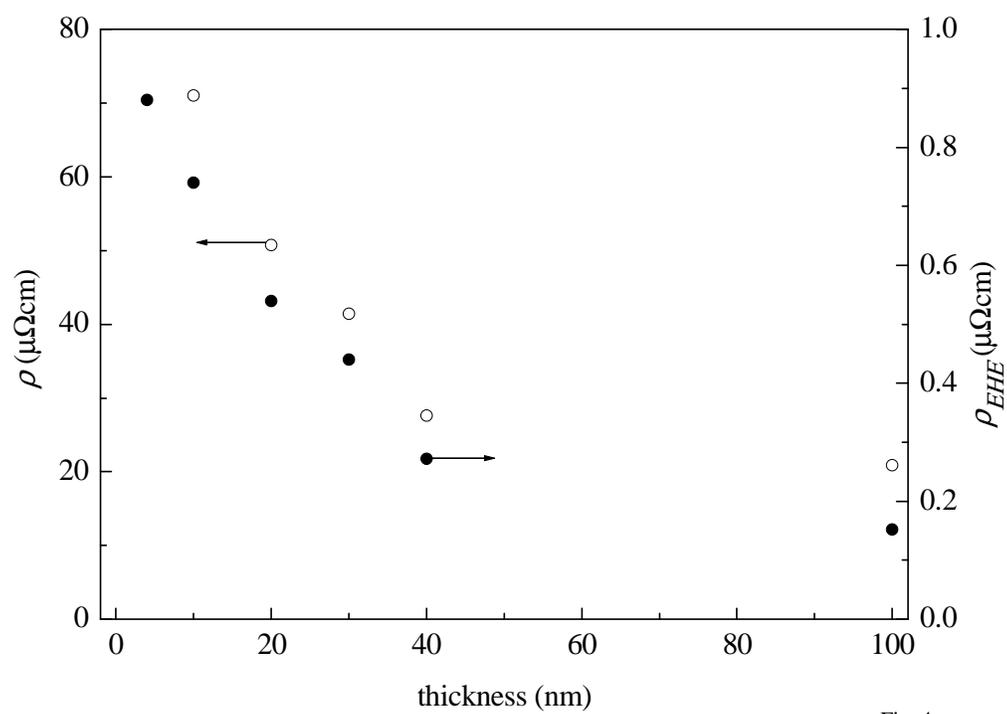

Fig. 4

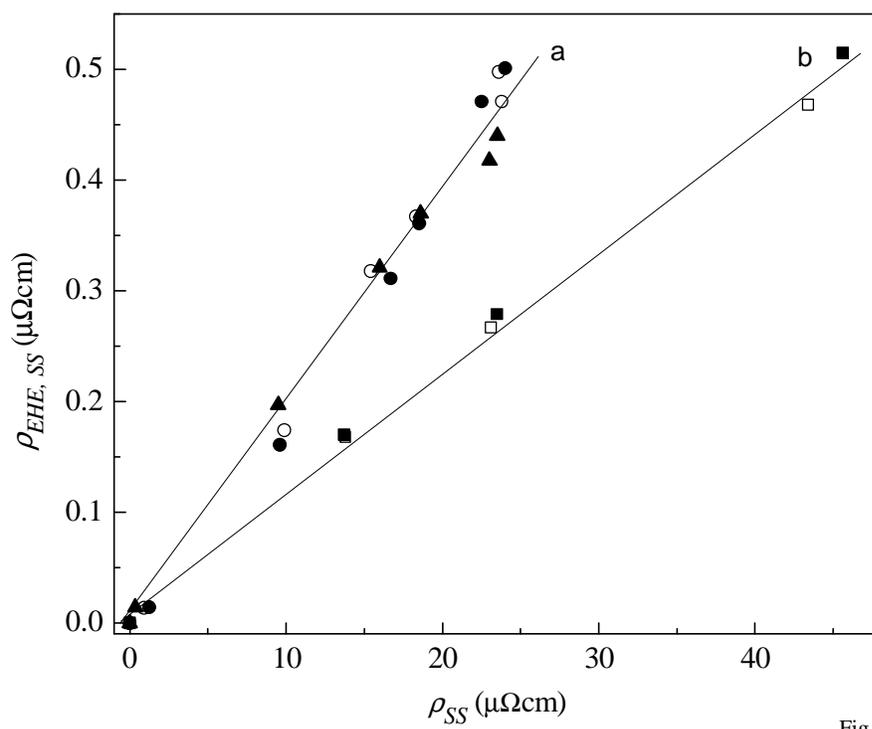

Fig.5

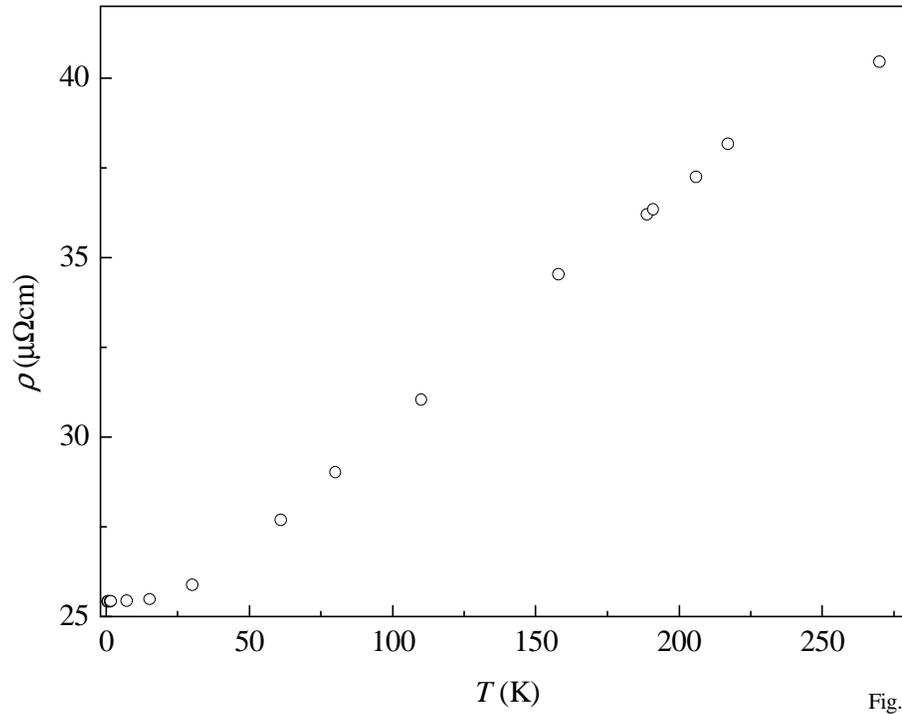

Fig. 6a

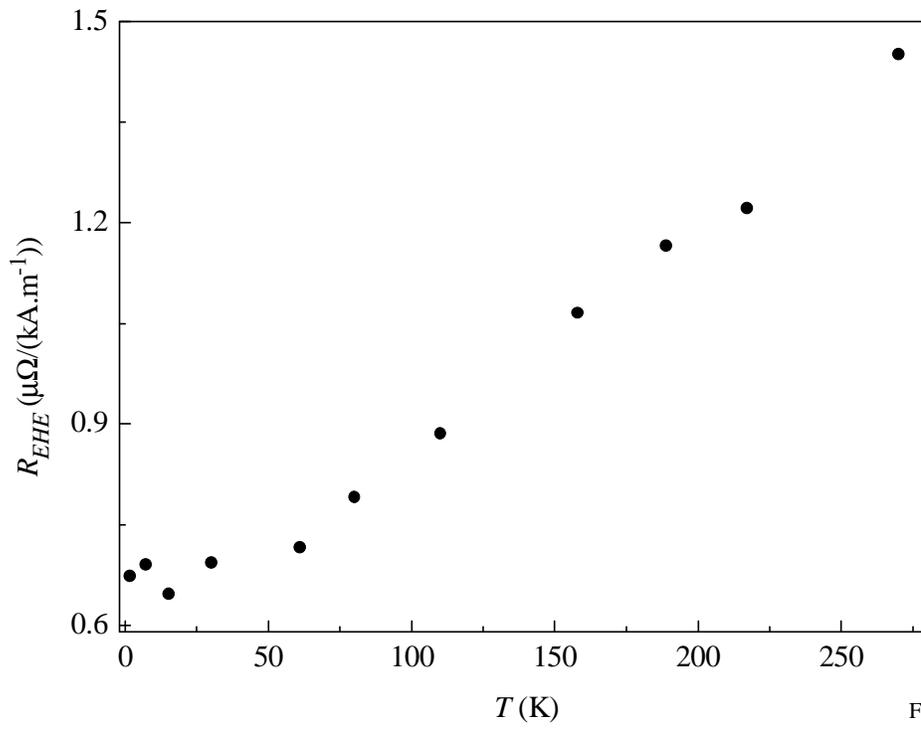

Fig. 6b

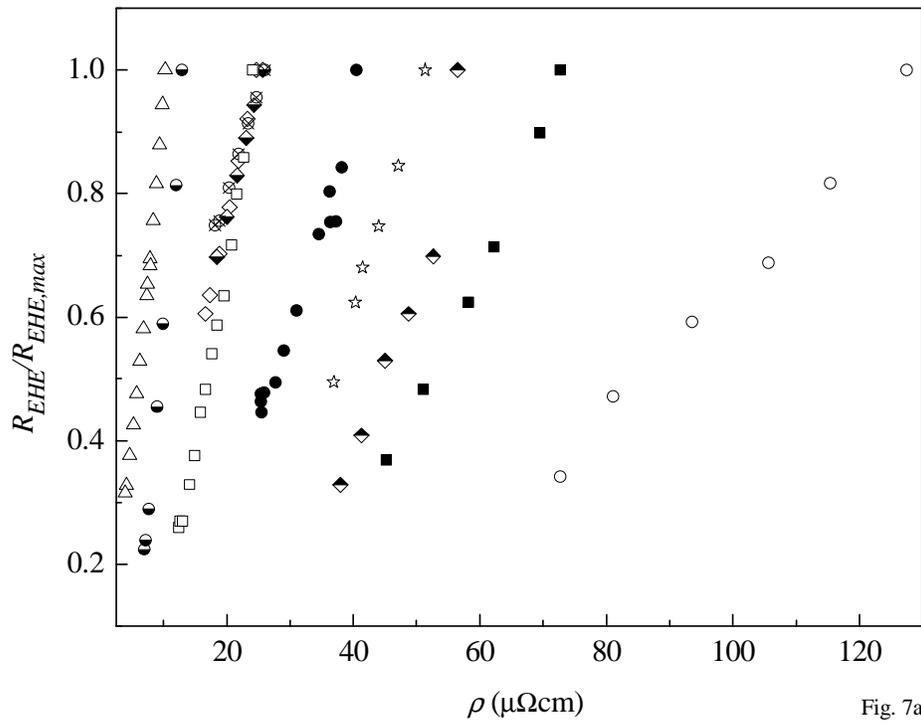

Fig. 7a

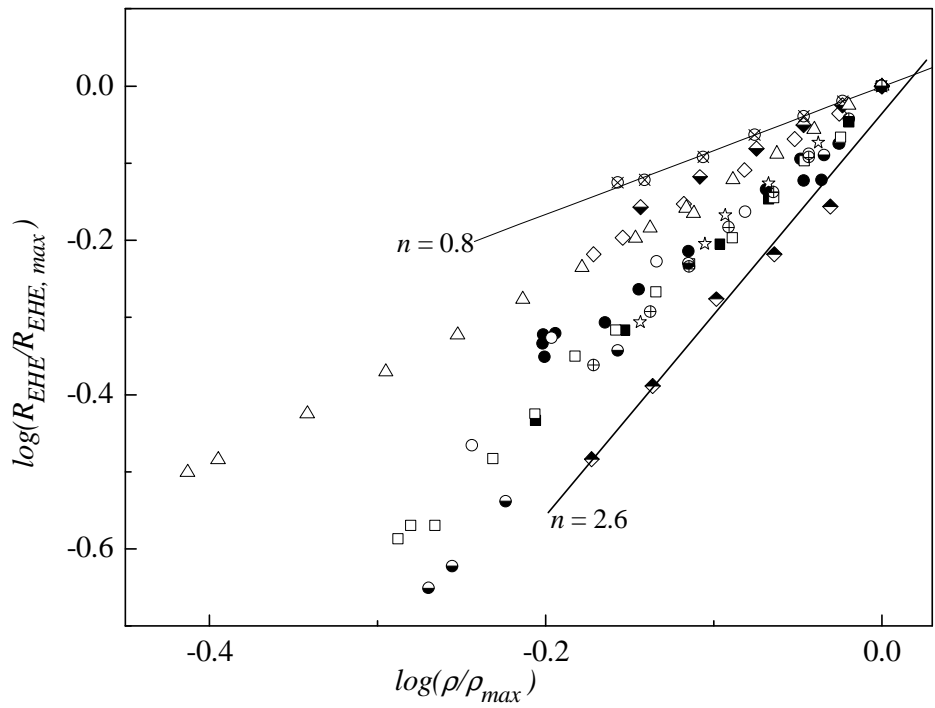

Fig. 7b

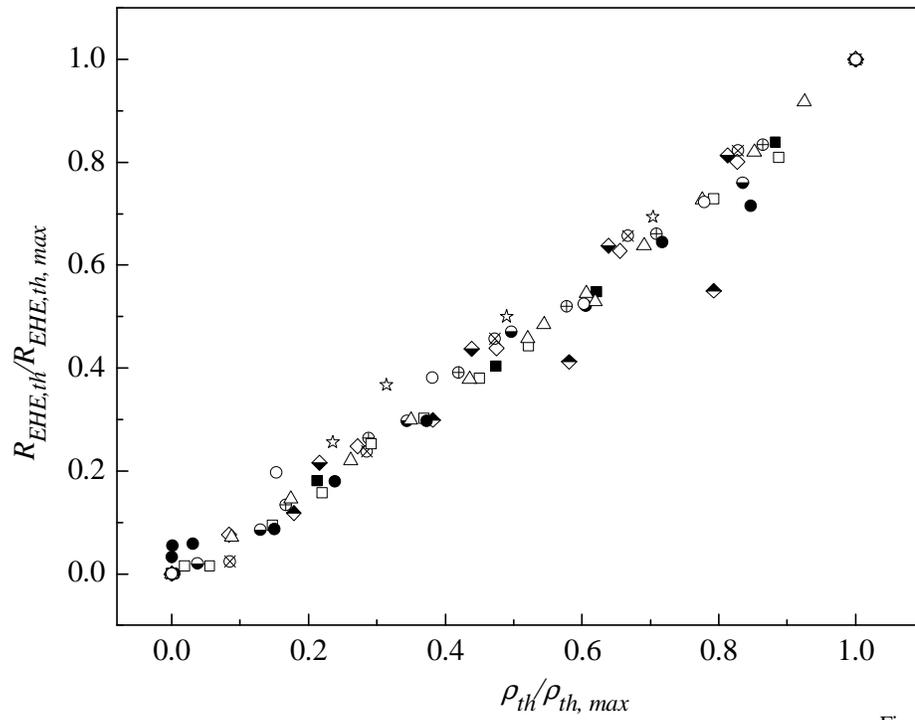

Fig.8